\documentclass[draftcls,10pt,onecolumn,english]{IEEEtran}
\usepackage[T1]{fontenc}
\usepackage[latin1]{inputenc}
\usepackage{textcomp}
\usepackage{amsmath}
\usepackage{amssymb}
\usepackage{tikz}
\usetikzlibrary{arrows}
\usetikzlibrary{fit}					
\usetikzlibrary{backgrounds}	
\makeatletter

\DeclareFontEncoding{LGR}{}{}
\DeclareTextSymbol{\~}{LGR}{126}
\newcommand{\lyxmathsym}[1]{\ifmmode\begingroup\def\b@ld{bold}
  \text{\ifx\math@version\b@ld\bfseries\fi#1}\endgroup\else#1\fi}

\makeatother

\usepackage{babel}
\newtheorem{theorem}{Theorem}
\newtheorem{lemma}{Lemma}

\newtheorem{corollary}{Corollary}
\newtheorem{definition}{Definition}

\newtheorem{remark}{Remark}

\def\tR{\tilde{R}}

\def\n{\nonumber\\}
\def\be{\begin{equation}}
\def\ee{\end{equation}}
\def\bes{\begin{equation*}}
\def\ees{\end{equation*}}
\def\beq{\begin{eqnarray}}
\def\eeq{\end{eqnarray}}
\def\beqs{\begin{eqnarray*}}
\def\eeqs{\end{eqnarray*}}

\def\ma{{\mathcal A}}
\def\mb{{\mathcal B}}

\def\mr{{\mathcal R}}
\def\ms{{\mathcal S}}

\def\mv{{\mathcal V}}

\def\mx{{\mathcal X}}
\def\my{{\mathcal Y}}

\def\e{\mathbb{E}}

\def\tv#1{\left\|#1\right\|_1}
\def\apx#1{\stackrel{#1}{\approx}}

\begin{document}

\title{Coordination via a Relay }

\author{Farzin Haddadpour, Mohammad Hossein Yassaee, Amin Gohari, Mohammad
Reza Aref\\

Information Systems and Security Lab (ISSL)\\
Department of Electrical Engineering, Sharif University of Technology,
Tehran, Iran\\
 Email: \{haddadpour,yassaee\}@ee.sharif.edu,\{aminzadeh,aref\}@sharif.edu }

\maketitle
\begin{abstract}
In this paper, we study the problem of coordinating two nodes which can only exchange information via a relay at limited rates. The nodes are allowed to do a two-round interactive two-way communication with the relay, after which they should be able to generate i.i.d.\ copies of two random variables with a given joint distribution within a vanishing total variation distance. 
We prove inner and outer bounds on the coordination capacity region for this problem. Our inner bound is proved using the technique of ``output statistics of random binning" that has recently been developed by Yassaee, et al.

\end{abstract}
\section{Introduction}
Coordination is the problem of producing dependent random variables over a network \cite{coordination}. This problem differs from traditional coding problems where the goal is to distribute explicit messages. The problem of coordination for a joint action has applications in distributed control and game theory \cite{Venkat,cuff}.
Two notions of coordination have been defined in \cite{coordination}, namely \emph{empirical coordination} and \emph{strong coordination}. 
In empirical coordination we want the empirical joint distribution of the actions to be close to the desired distribution, whereas in the strong coordination we want the total variation 
distance between the joint probability distribution of the actions, and the i.i.d. copies of the given distribution to be negligibly small. 
In other words, the generated distribution and the i.i.d.\ distribution should be statistically indistinguishable. These are two different notions of coordination. 
In this paper we study the strong notion of coordination. 

As discussed in \cite{coordination}, nodes in a network can cooperate arbitrarily without any communication if they are provided with sufficient common randomness.
However \cite{coordination} argues that problem becomes nontrivial if the action of some of the nodes is specified by nature. 
We believe that this is not the \emph{only} situation where the problem becomes nontrivial. Suppose that two nodes of a network want to cooperate with each other while remaining anonymous to each other. They can obtain anonymity through a proxy (relay) who privately exchanges messages with the two nodes. 
Since the two nodes cannot directly talk to each other, they will not be able to directly share randomness.
However they may attempt to create common randomness indirectly through the relay. But the rate of this common randomness will be bounded from above by the communication rate constraints between 
the nodes and the relay. Furthermore creating common randomness for later use may not be the optimal strategy if the final goal is coordination. The communication links between the nodes and the relay are rate limited, and hence there may exist more economic ways of  using this resource. Inspired by this discussion, we propose the following model as an attempt to understand the use of a relay in cooperation of two nodes whose actions are \emph{not} specified by nature.
\begin{figure}
\centering
\tikzstyle{background}=[rectangle,
                                                fill=yellow!20,
                                                inner sep=0.2cm,
                                                rounded corners=5mm,
                                               scale=1 ]
\begin{tikzpicture}[>=triangle 45]
\draw (0,0) node (a){$A_1$} (4,0) node (b){$A_0$} (8,0) node (c){$A_2$};
\draw (0,-1.5) node (a1){$A_1$} (4,-1.5) node (b1){$A_0$} (8,-1.5) node (c1){$A_2$};
\draw[->] (a) to[out=30,in=150] node [sloped,above](d){$R_{b_1}$}(b);
\draw[<-] (b) to[out=30,in=150] node [sloped,above]{$R_{b_2}$}(c);

\draw[<-](a1) to[out=-30,in=-150]node [sloped,below](d1){$R_{f_1}$} (b1);
\draw[->] (b1) to[out=-30,in=-150] node [sloped,below]{$R_{f_2}$}(c1);
\begin{pgfonlayer}{background}
        \node [background,
                    fit=(a)(d)(c),
                    ] {};
    \end{pgfonlayer}
    
    \begin{pgfonlayer}{background}
        \node [background,
                    fit=(a1)(d1)(c1),
                    ] {};
    \end{pgfonlayer}

\end{tikzpicture}
\caption{The model for coordination via a relay. In the first step nodes $A_1$ and $A_2$ communicate to the relay node $A_0$ (as in the top subfigure). In the second step
the relay communicates to $A_1$ and $A_2$ (as in the bottom subfigure). }\label{fig:1}
\vskip -0.5cm
\end{figure}
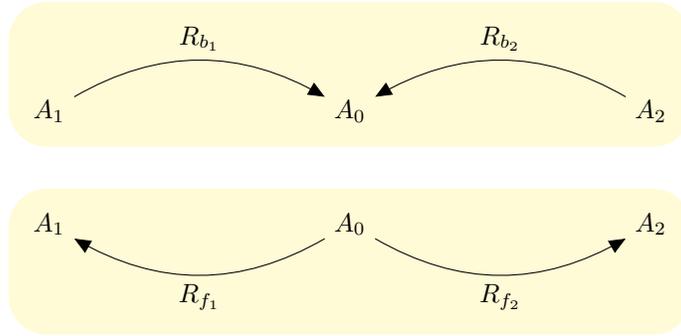

As shown in Fig. \ref{fig:1}, we assume that there are four links between the relay ($A_0$) and the two nodes ($A_1$ and $A_2$). The noiseless forward links from the relay to the the first and second nodes have rates $R_{f_1}$ and $R_{f_2}$ respectively. The backward links have rates $R_{b_1}$ and $R_{b_2}$. As can be seen from the figure, the nodes use the backward links first to communicate to the relay, after which the relay communicates back to the nodes using the forward links. The goal of the two parties is to generate i.i.d.\ copies of $Y_1$ and $Y_2$ jointly distributed according to a given $p(y_1, y_2)$ within a vanishing total variation distance. 
We don't assume any common randomness shared between $A_1$ and $A_2$ since the two nodes don't share any resources beyond private communication links with the proxy. However, private randomization is allowed at all the three nodes.
Further we could have added a separate rate limited \emph{public} forward link from the proxy to all the nodes, where all the bits put on this link will become available to all the parties. 
Adding this link would make our model to resemble the model proposed by Wyner \cite{Wyner} where a set of random bits were being simultaneously transmitted  to two parties. However, 
we have excluded this from our model for simplicity. 

Since the two nodes are initially communicating at rates $R_{b_1}$ and $R_{b_2}$, the nodes can use these only to generate pairwise common randomness between themselves and the proxy. Thus one can reinterpret the model as a one-way communication problem from the relay to the two nodes in the presence of pairwise common randomness. This has been the motivation for naming $R_{f_1}$ and $R_{f_2}$ as forward links although they are being used in the second step of the protocol.

It is noteworthy  that to see when $R_{f_{1}}=0$ and $R_{b_{1}}=\infty$ our model reduces to the one considered by Cuff in \cite{cuff}. If $R_{f_{1}}=0$,
the first node does not receive any feedback and has to create the i.i.d.\ copies of $Y_1^n$ by itself. Since $R_{b_{1}}=\infty$, the first node can send $Y_1^n$ completely to the relay. The relay is receiving $R_{b_{2}}$ bits from the second node which can be understood as a common randomness shared between $A_0$ and $A_2$. Thus, our problem reduces to the problem of \cite{cuff}. If $R_{f_{1}}=\infty$, the problem reduces to a special case of the problem studied in \cite{aminzade}. In this case the relay is effectively coordinating with the second node because the relay can send its reconstruction of $Y_1^n$ to the first node using the forward link $R_{f_{1}}$ of infinite capacity. Thus this would be the problem of generating $Y_1^n$ and $Y_2^n$ using a two-round communication scheme when the two node share no common randomness. When $R_{b_{1}}=R_{b_{2}}=\infty$, the problem reduces to that of coordinating $A_1$ and $A_2$ when there are pairwise common randomness shared between $(A_0, A_1)$ and $(A_0, A_2)$ but no common randomness shared among the three. Finally when $R_{b_{1}}=R_{b_{2}}=0$ the problem
reduces to a problem that resembles Wyner's model \cite{Wyner}.

We prove an inner and an outer bound on our model. We show that the inner and the outer bound match in certain special cases, two of which are of special interest: one is when 
$R_{b_{1}}=R_{b_{2}}=\infty$, i.e. an infinite pairwise common randomness, the other is when $R_{b_{1}}=R_{b_{2}}=0$, i.e. no pairwise common randomness. We show that when $R_{b_{1}}=R_{b_{2}}=\infty$, the capacity region is the one where $R_{f_1}+R_{f_2}$ is greater than or equal to the mutual information between $Y_1$ and $Y_2$. In the other extreme case both $R_{f_1}$ and $R_{f_2}$ have to be larger than Wyner's common information. This provides insights on the role of \emph{pairwise} common randomness. 

This paper is organized as follows: in Section II, we introduce the
basic notations and definitions used in this paper. Section III contains
the main results of the paper, and Section IV and V includes the proofs.

\section{Definitions}

\subsection{Notation}
 In this paper, we use $p^U_{\ma}$ to denote the uniform distribution over the set $\ma$ and $p(x^n)$ to denote the i.i.d. pmf $\prod_{i=1}^np(x_i)$, unless otherwise stated.
Also we use $X_{\ms}$ to denote $(X_j:j\in\ms)$.
 The total variation between two pmf's $p$ and $q$ on the same alphabet $\mx$ , is denoted by $\tv{p(x)-q(x)}$. When a pmf itself is random, we use capital letter, e.g. $P_X$. 
\begin{remark} Similar to \cite{cuff} in this work we frequently use the concept of \emph{random} pmfs, which we denote by capital letters (e.g. $P_X$). For any countable set $\mx$ let $\Delta^{\mx}$ be the probability simplex for distributions on $\mx$. A random pmf $P_X$ is a probability distribution over $\Delta^{\mx}$. In other words, if we use $\Omega$ to denote the sample space, the mapping $\omega\in \Omega \mapsto P_X(x;\omega)$ is a random variable for all $x\in\mx$ such that $P_X(x;\omega)\geq 0$ and $\sum_{x}P_X(x;\omega)=1$ for all $\omega$. Thus, $\omega\mapsto P_X(\cdot;\omega)$ is a vector of random variables, which we denote by $P_X$. We can definite $P_{X,Y}$ on product set $\mx\times\my$ in a similar way. We note that we can continue to use the law of total probability with random pmfs (e.g. to write $P_X(x)=\sum_{y}P_{XY}(x,y)$ meaning that $P_X(x;\omega)=\sum_yP_{XY}(x,y;\omega)$ for all $\omega$) and the conditional probability pmfs (e.g. to write $P_{Y|X}(y|x)=\frac{P_{XY}(x,y)}{P_X(x)}$ meaning that $P_{Y|X}(y|x;\omega)=\frac{P_{XY}(x,y;\omega)}{P_X(x;\omega)}$ for all $\omega$).
\end{remark}

\subsection{Problem Statement}

Consider the problem of strong coordination over a network with a relay, as depicted in Figure \ref{fig:1}. 
In this setting, there are three nodes $A_1$, $A_0$ and $A_2$. They do not share any common randomness, but private randomization is allowed. Let $M_i$ be the private randomness at node $A_i$. A $(n, R_{f_{1}}, R_{b_{1}}, R_{f_{2}}, R_{b_{2}})$ \emph{coordination code} consists of
\begin{itemize}
\item Two encoders at nodes $A_k, k=1,2$, that map $\mathcal{M}_k$ to $[1:2^{nR_{b_k}}]$.
\item Two encoders at the relay node $A_0$, that map $\mathcal{M}_0\times \mathcal{B}_1\times \mathcal{B}_2$ to $[1:2^{nR_{f_k}}]$ for $k=1,2$.
\item Two decoders at nodes $A_k, k=1,2$, that map $\mathcal{M}_k\times \mathcal{B}_k\times \mathcal{F}_k$ to $\mathcal{Y}_k^n$.
\end{itemize}

\begin{definition} A joint distribution $q(y_{1},y_{2})$ is said to be in the admissible region of 
the rate tuple $(R_{f_{1}}, R_{b_{1}}, R_{f_{2}}, R_{b_{2}})$ if one can find a sequence of $(n, R_{f_{1}}, R_{b_{1}}, R_{f_{2}}, R_{b_{2}})$
coordination codes for $n=1,2,...$ whose induced joint distributions have marginal distributions $p(y_{1}^{n},y_{2}^{n})$
that satisfy
\[
\lim_{n\rightarrow\infty}\tv{p(y_{1}^{n},y_{2}^{n})\text{\textminus}\prod_{i=1}^{n}q(y_{1,i},y_{2,i})}=0.
\]
\end{definition}

\begin{definition} Given a joint distribution $q(y_{1},y_{2})$, the coordination rate region is the closure of the set of rate tuples $(R_{f_{1}},R_{b_{1}},R_{f_{2}},R_{b_{2}})$ that admit the channel $q(y_{1},y_{2})$.
\end{definition}

\section{Main Results}

\begin{theorem}[Inner bound]
The following region forms an inner bound to the coordination rate region for $q(y_1,y_2)$: $\mathcal{R}_{\mathsf{in}}$ is  the set of all non-negative rate tuples $(R_{f_1},R_{b_1},R_{f_2},R_{b_2})$, for which there exists $p(u,v,w,y_1,y_2)\in T_{\mathsf{in}}$ such that
\begin{align}
R_{b_{1}}+R_{f_{1}}+R_{b_{2}}+R_{f_{2}}&\ge I(Y_{1}Y_{2};VUW)+I(U;V|W)+I(W;Y_{1}Y_{2}),\nonumber\\
R_{b_{1}}+R_{f_{1}}&\ge I(Y_{1}Y_{2};VW),\nonumber\\
R_{b2}+R_{f2}&\ge I(Y_{1}Y_{2};UW),\nonumber\\
R_{f_{2}}+R_{f_{1}}&{\ge}I(U;V|W)+I(W;Y_{1}Y_{2}), \label{eq:c0}
\end{align}
where
\begin{align*}
T_{\mathsf{in}}=\{p(u,v,w,y_1,y_2):&(Y_1,Y_2)\sim q(y_1,y_2),\\
&Y_2-UW-VW-Y_1\}.\end{align*}

\end{theorem}
\begin{theorem}[Outer bound]
Take a desired distribution $q(y_1,y_2)$. Then the coordination rate region is contained in the region $\mathcal{R}_{\mathsf{out}}$ which is the closure of the set of all non-negative rate tuples $(R_{f_1},R_{b_1},R_{f_2},R_{b_2})$, for which there exists $p(u,v,y_1,y_2)\in T_{\mathsf{out}}$ such that
\begin{equation}
\begin{split}
R_{b_{1}}+R_{f_{1}}&\ge I(Y_{1}Y_{2};V),\\
R_{b_2}+R_{f_2}&\ge I(Y_{1}Y_{2};U),\\
R_{f_{2}}+R_{f_{1}}&\ge\max\{I(U;Y_{1}),I(V;Y_{2})\},
\end{split}
\end{equation}
where
\begin{align*}
T_{\mathsf{out}}=\{p(u,v,y_1,y_2):&(Y_1,Y_2)\sim q(y_1,y_2),\\
&\qquad Y_2-U-Y_1,\\&\qquad Y_2-V-Y_1,\\&
|\mathcal{U}|\leq |\my_1|\times|\my_2|+1,
\\&
|\mv|\leq |\my_1|\times|\my_2|+1
\}.\end{align*}
\end{theorem}
\begin{corollary} The inner bound and the outer bound match when $R_{b_1}=R_{b_2}=\infty$, both reducing to $R_{f_{2}}+R_{f_{1}}\ge I(Y_1;Y_2)$. This corresponds to the case of infinite pairwise common randomness and has not been considered (to best of our knowledge) in the previous works. When $R_{f_1}=\infty$, the inner and outer bound reduce to $R_{b_{2}}+R_{f_{2}}$ being greater than or equal to Wyner's common information. The inner and outer bound also match when $R_{f_1}=0$ and $R_{b_1}=\infty$. To see this let $V=Y_1$ and $W=cont.$ in the inner bound. On the other hand the optimal choice for $V$ in the outer bound is $V=Y_1$. Thus both regions reduce to the following region that matches the one given in \cite{cuff}.
\begin{equation*}
\begin{split}
R_{b_2}+R_{f_2}&\ge I(Y_{1}Y_{2};U),\\
R_{f_{2}}&\ge I(U;Y_{1}).
\end{split}
\end{equation*}
Another extreme case is when $R_{b_1}=R_{b_2}=0$. Here we take $U=V=cont.$ in the inner bound. It is easy to see that both the inner and outer bound reduce to $R_{f_1}$ and $R_{f_2}$ being greater than or equal to Wyner's common information. Comparing this case with Wyner's model, we see that an optimal strategy is to send the same message to both $A_1$ and $A_2$ (which is expected when $R_{b_1}=R_{b_2}=0$). The inner and outer bound also match when $Y_{1}=(A, B)$, $Y_{2}=(A, C)$ for mutually independent random variable $A$, $B$ and $C$. 
\end{corollary}


\section{Achievability}
We apply the techniques of \cite{me} to prove the achievability of the theorem. We begin by a providing a summary of the lemmas we need. In the following subsection we provide the proof.
\subsection{Review of probability approximation via random binning \cite{me}}
Let $(X_{[1:T]}, Y)$ be a DMCS distributed according to a joint pmf $p_{X_{[1:T]},Y}$ on a countably infinite set $\prod_{i=1}^T \mx_i\times \my$. A distributed  random  binning consists of a set of random mappings $\mb_i: \mx_i^n\rightarrow [1:2^{nR_i}]$, $i\in[1:T]$, in which $\mb_i$ maps each sequence of $\mx_i^n$ uniformly and independently to $[1:2^{nR_i}]$. We denote the random variable $\mb_t(X_t^n)$ by $B_t$. A random distributed  binning induces the following \emph{random pmf} on the set $\mx_{[1:T]}^n\times\my^n\times\prod_{t=1}^T [1:2^{nR_t}]$,
\[
P(x^n_{[1:T]},y^n,b_{[1:T]})=p(x_{[1:T]}^n,y^n)\prod_{t=1}^T\mathbf{1}\{\mb_t(x_t^n)=b_t\}.
\]
\begin{theorem}[\cite{me}]
\label{thm:re}
If for each $\ms\subseteq [1:T]$, the following constraint holds 
\be
\sum_{t\in\ms}R_t<H(X_{\ms}|Y),
\ee
then as $n$ goes to infinity, we have
\be
\e\tv{P(y^n,b_{[1:T]})-p(y^n)\prod_{t=1}^T p^U_{[1:2^{nR_t}]}(b_t)}\rightarrow 0.
\ee
\end{theorem}
\par
We now consider another region for which we can approximate a specified pmf. This region is the Slepian-Wolf region for reconstructing $X^n_{[1:T]}$ in the presence of $(B_{1:T}, Y^n)$ at the decoder. 
 As in the achievability proof of the \cite[Theorem 15.4.1]{cover:book}, we can define a decoder with respect to any fixed distributed binning. We denote the decoder by the random conditional pmf $P^{SW}(\hat{x}^n_{[1:T]}|y^n,b_{[1:T]})$ (note that since the decoder is a function, this pmf takes only two values, 0 and 1). Now we write the Slepian-Wolf theorem in the following equivalent form. See \cite{me} for details.
\begin{lemma}\label{le:sw}
If for each $\ms\subseteq [1:T]$, the following constraint holds 
\be
\sum_{t\in\ms}R_t>H(X_{\ms}|X_{\ms^c}, Y),
\ee
then as $n$ goes to infinity, we have
\bes
\e\tv{P(x^n_{[1:T]},y^n,\hat{x}^n_{[1:T]})-p(x^n_{[1:T]},y^n)\mathbf{1}\{\hat{x}^n_{[1:T]}=x^n_{[1:T]}\}}\rightarrow 0.\ees
\end{lemma}  
\begin{definition}\label{def:1}
For any random pmfs $P_X$ and $Q_X$ on $\mx$, we say $P_X\stackrel{\epsilon}{\approx}Q_X$ if $\e\tv{P_X-Q_X}<\epsilon$. Similarly we use $p_X\apx{\epsilon}q_x$ for two (non-random) pmfs to denote the total variation constraint $\tv{p_X-q_X}<\epsilon$.\end{definition}

\begin{lemma}\label{le:total}We have
\begin{enumerate}
\item
$\tv{p_Xp_{Y|X}-q_{X}p_{Y|X}}=\tv{p_X-q_X}$\\
$~~~~~~~~~~~\quad\tv{p_X-q_X}\le\tv{p_Xp_{Y|X}-q_{X}q_{Y|X}}$
\item If $p_Xp_{Y|X}\stackrel{\epsilon}{\approx}q_Xq_{Y|X}$, then there exists $x\in\mx$ such that $p_{Y|X=x}\stackrel{2\epsilon}{\approx}q_{Y|X=x}$.
\item If $P_X\stackrel{\epsilon}{\approx}Q_X$ and  $P_XP_{Y|X}\stackrel{\delta}{\approx}P_XQ_{Y|X}$, then $P_{X}P_{Y|X}\stackrel{\epsilon+\delta}{\approx}Q_{X}Q_{Y|X}$.
\end{enumerate}
\end{lemma}
\subsection{Proof of Theorem 1}
The proof is divided into three parts. In the first part we introduce two protocols each of which induces a pmf on a certain set of r.v.'s. The first protocol has the desired i.i.d.\ property on $Y_1^n$ and $Y_2^n$, but leads to no concrete coding algorithm. However the second protocol is suitable for construction of a code, with one exception: the second protocol is assisted with an extra common randomness that does not really exist in the model. In the second part we find conditions on $R_{b_1}, R_{b_2}, R_{f_1}, R_{f_2}$ implying that these two induced distributions are almost identical. In the third part of the proof, we eliminate the extra common randomness given to the second protocol without disturbing the pmf induced on the desired random variables ($Y_1^n$ and $Y_2^n$) significantly. This makes the second protocol useful for code construction.

\emph{Part (1) of the proof:} 
We define two protocols each of which induces a joint distribution on random variables that are defined during the protocol. 

\emph{Protocol A. }
Let $(W^n,U^n,V^n,Y_1^n,Y_2^n)$ be i.i.d. and distributed according to $p(w,v,u,y_1,y_2)$ such that the marginal pmf of $(Y_1,Y_2)$ satisfies $p(y_1,y_2)=q(y_1,y_2)$. 
Consider the following random binning:
\begin{itemize}
\item To each sequence $w^n$, assign a random bin index $g_0\in[1:2^{n\tR_0}]$.
\item To each pair $(w^n,v^n)$, assign three random bin indices $g_1\in[1:2^{n\tR_1}]$, $b_1\in[1:2^{nR_{b_1}}]$ and $f_1\in[1:2^{nR_{f_1}}]$.
\item To each pair $(w^n,u^n)$, assign three random bin indices $g_2\in[1:2^{n\tR_2}]$, $b_2\in[1:2^{nR_{b_2}}]$ and $f_2\in[1:2^{nR_{f_2}}]$.
\item We use a Slepian-Wolf decoder to recover $\hat{w}_1^n,\hat{v}^n$ from $(g_0,g_1,b_1,f_1)$, and another Slepian-Wolf decoder to recover $\hat{w}_2^n,\hat{u}^n$ from $(g_0,g_2,b_2,f_2)$.  The rate constraints for the success of these decoders will be imposed later, although these decoders can be conceived even when there is no guarantee of success. 
\end{itemize}
The random\footnote{The pmf is random because we are doing a random binning assignment in the protocol.} pmf induced by the random binning, denoted by $P$, can be expressed as follows:
\begin{align*}
&P(g_{0}|w^n)P(g_{1}b_1f_1|w^nv^n)P(g_{2}b_2f_2|w^nu^n)p(w^n, v^n, u^n)
~\times\\& P^{SW}(\hat{w}_1^n,\hat{v}^n|g_0,g_1,b_1,f_1)P^{SW}({\hat{w}_2}^n,\hat{u}^n|g_0,g_2,b_2,f_2)\times\\&
\quad p( y_{1}^n|w^nu^n)p(y_{2}^n|w^nv^n).
\end{align*}

\emph{Protocol B.} In this protocol we assume that the nodes have access to the extra common randomness $(G_0, G_1, G_2)$ where $G_0,G_1,G_2$ are mutually independent random variables distributed uniformly over the sets $[1:2^{n\tR_0}],\ [1:2^{n\tR_1}]$ and $[1:2^{n\tR_2}]$, respectively. Now, we use the following protocol:
\begin{itemize}
\item At the first stage, the node $A_1$ chooses an index $b_1\in[1:2^{nR_{b_1}}]$ uniformly at random and sends it to the node $A_0$. Also the node $A_2$ independently chooses an index $b_2\in[1:2^{nR_{b_2}}]$ uniformly at random and sends it to the node $A_0$.
\item In the second stage, knowing $(g_0,g_1,g_2,b_1,b_2)$, the node $A_0$ generates sequences $(w^n,v^n,u^n)$ according to  the conditional pmf $P(w^n,v^n,u^n|g_0,g_1,g_2,b_1,b_2)$ of the protocol A. Then it sends the bin indices $f_1(w^n,v^n)$ and $f_2(w^n,u^n)$ to the nodes $A_1$ and $A_2$, respectively.
\item At the final stage, the node $A_1$, knowing $(g_0,g_1,b_1,f_1)$ uses the Slepian-Wolf decoder $P^{SW}(\hat{w}_1^n,\hat{v}^n|g_0,g_1,b_1,f_1)$ to obtain an estimate of $(w^n,v^n)$. Then, it generates a sequence $y_1^n$ according to $p_{Y^n|W^nV^n}(y_1^n|\hat{w}_1^n,\hat{v}^n)$. The node $A_2$ proceeds in a similar way.
\end{itemize}
The random pmf induced by the protocol, denoted by $\hat{P}$, factors as
\begin{align}
&p^U(g_{[0:2]})p^U(b_1)p^U(b_2)P(w^n,v^n,u^n,f_{[1:2]}|g_{[0:2]}b_{[1:2]})\times\n
&P^{SW}(\hat{w}_1^n,\hat{v}^n|g_0,g_1,b_1,f_1)P^{SW}({\hat{w}_2}^n,\hat{u}^n|g_0,g_2,b_2,f_2)\times\n
&\qquad p(y_1^n|\hat{w}_1^n,\hat{v}^n)p(y_2^n|\hat{w}_2^n,\hat{u}^n)\label{eq:pmf}
\end{align}

\emph{Part (2) of the proof: Sufficient conditions that make the induced pmfs approximately the same}: To find the constraints that imply that the pmf $\hat{P}$ is close to the pmf $P$ in total variation distance, 
we start with $P$ and make it close to $\hat{P}$ in a few steps. The first step is to observe that $g_0$, $(g_1,b_1)$ and $(g_2,b_2)$ are the bin indices of $w^n$, $(w^n,v^n)$ and $(w^n,u^n)$, respectively. Substituting $T=3$, $X_1=W$, $X_2=WV$, $X_3=WU$ and $Y=\emptyset$ in Theorem \ref{thm:re}, implies that if
\be\label{eq:c1}\begin{split}
\tR_0&<H(W),\\
\tR_0+\tR_1+R_{b_1}&<H(WV),\\
\tR_0+\tR_2+R_{b_2}&<H(WU),\\
\tR_0+\tR_1+\tR_2+R_{b_1}+R_{b_2}&<H(WVU),
\end{split}\ee
then there exists $\epsilon_0^{(n)}\rightarrow 0$ such that $P(g_{[0:2]},b_1,b_2)\apx{\epsilon_0^{(n)}}p^U(g_{[0:2]})p^U(b_1)p^U(b_2)=\hat{P}(g_{[0:2]},b_1,b_2)$. 
This implies
\begin{align}
&\hat{P}(g_{[0:2]},b_1,b_2,w^n,v^n,u^n,\hat{w}_1^n,\hat{v^n},{\hat{w}}_2^n,\hat{u}^n)\apx{\epsilon_0^{(n)}}
P(g_{[0:2]},b_1,b_2,w^n,v^n,u^n,\hat{w}_1^n,\hat{v^n},{\hat{w}}_2^n,\hat{u}^n)\label{eqn:ee1}
\end{align}

The next step is to see that for the Slepian-Wolf decoders of the first protocol to work well, Lemma \ref{le:sw} requires imposing the following constraints:
\be\label{eq:c2}\begin{split}
\tR_1+R_{b_1}+R_{f_1}&\ge H(V|W),\\
\tR_0+\tR_1+R_{b_1}+R_{f_1}&\ge H(WV),\\
\tR_2+R_{b_2}+R_{f_2}&\ge H(U|W),\\
\tR_0+\tR_2+R_{b_2}+R_{f_2}&\ge H(WU),
\end{split}\ee
then for some vanishing sequence $\epsilon_1^{(n)}$, we have
\begin{align*}
P(&g_{[0:2]},b_1,b_2,w^n,v^n,u^n,\hat{w}_1^n,\hat{v^n},{\hat{w}}_2^n,\hat{u}^n)\\&\apx{\epsilon_1^{(n)}}P(g_{[0:2]},b_1,b_2,w^n,v^n,u^n)\mathbf{1}\{\hat{w}_1^n=w^n,\hat{v^n}=v^n,{\hat{w}_2}^n=w^n,\hat{u}^n=u^n\}.
\end{align*}
Using equation \eqref{eqn:ee1} we have 
\begin{align*}
\hat{P}(&g_{[0:2]},b_1,b_2,w^n,v^n,u^n,\hat{w}_1^n,\hat{v^n},{\hat{w}}_2^n,\hat{u}^n)\\&\apx{\epsilon_0^{(n)}+\epsilon_1^{(n)}}P(g_{[0:2]},b_1,b_2,w^n,v^n,u^n)\mathbf{1}\{\hat{w}_1^n=w^n,\hat{v^n}=v^n,{\hat{w}_2}^n=w^n,\hat{u}^n=u^n\}.
\end{align*}
The third part of Lemma \ref{le:total} implies that
\begin{align*}
\hat{P}(&g_{[0:2]},b_1,b_2,w^n,v^n,u^n,\hat{w}_1^n,\hat{v^n},{\hat{w}}_2^n,\hat{u}^n)
 p(y_1^n|\hat{w}_1^n,\hat{v}^n)p(y_2^n|\hat{w}_2^n,\hat{u}^n)\\&\apx{\epsilon_0^{(n)}+\epsilon_1^{(n)}}P(g_{[0:2]},b_1,b_2,w^n,v^n,u^n)\mathbf{1}\{\hat{w}_1^n=w^n,\hat{v^n}=v^n,{\hat{w}_2}^n=w^n,\hat{u}^n=u^n\}
 p(y_1^n|\hat{w}_1^n,\hat{v}^n)p(y_2^n|\hat{w}_2^n,\hat{u}^n)\\
&~~~=~~~~P(g_{[0:2]},b_1,b_2,w^n,v^n,u^n)\mathbf{1}\{\hat{w}_1^n=w^n,\hat{v^n}=v^n,{\hat{w}_2}^n=w^n,\hat{u}^n=u^n\} p(y_1^n|w_1^n,v^n)p(y_2^n|w_2^n,u^n).
\end{align*}
Thus,
\begin{align*}
\hat{P}(&g_{[0:2]},b_1,b_2,w^n,v^n,u^n,\hat{w}_1^n,\hat{v^n},{\hat{w}}_2^n,\hat{u}^n, y_1^n, y_2^n)\\&\apx{\epsilon_0^{(n)}+\epsilon_1^{(n)}}P(g_{[0:2]},b_1,b_2,w^n,v^n,u^n,y_1^n,y_2^n)\mathbf{1}\{\hat{w}_1^n=w^n,\hat{v^n}=v^n,{\hat{w}_2}^n=w^n,\hat{u}^n=u^n\}.
\end{align*}
Using the second item in part 1 of Lemma \ref{le:total} we conclude that
\begin{align*}
\hat{P}(&g_{[0:2]},y_1^n, y_2^n)\apx{\epsilon_0^{(n)}+\epsilon_1^{(n)}}P(g_{[0:2]},y_1^n,y_2^n).\end{align*}
 In particular, the marginal pmf of $(Y_1^n,Y_2^n)$ of the RHS of this expression is equal to $p(y_1^n,y_2^n)$ which is the desired pmf.

\emph{Part (3) of the proof:} 
 In the protocol we assumed that the nodes have access to an external randomness $G_{[0:2]}$ which is not present in the model. Nevertheless, we can assume that the nodes agree on an instance $g_{[0:2]}$ of $G_{[0:2]}$. In this case, the induced pmf $\hat{P}(y_1^n,y_2^n)$ changes to the conditional pmf $\hat{P}(y_1^n,y_2^n|g_{[0:2]})$. But if $G_{[0:2]}$ is independent of $(Y_1^n,Y_2^n)$, then the conditional pmf $\hat{P}(y_1^n,y_2^n|g_{[0:2]})$ is also close to the desired distribution. To obtain the independence, we again use Theorem \ref{thm:re}. Substituting $T=3$, $X_1=W$, $X_2=WV$, $X_3=WU$ and $Y=Y_1Y_2$ in Theorem \ref{thm:re}, asserts that if
\be\label{eq:c3}\begin{split}
\tR_0&<H(W|Y_1Y_2),\\
\tR_0+\tR_1&<H(WV|Y_1Y_2),\\
\tR_0+\tR_2&<H(WU|Y_1Y_2),\\
\tR_0+\tR_1+\tR_2&<H(WVU|Y_1Y_2),
\end{split}\ee
then $P(y_1^n,y_2^n,g_{[0:2]})\apx{\epsilon_2^{(n)}}p^U(g_{[0:2]})p(y_1^n,y_2^n)$, for some vanishing sequence $\epsilon_2^{(n)}$. Using triangular inequality for total variation, we have $\hat{P}(y_1^n,y_2^n,g_{[0:2]})\apx{\epsilon^{(n)}}p^U(g_{[0:2]})p(y_1^n,y_2^n)$, where $\epsilon^{(n)}=\sum_{i=0}^2\epsilon_i^{(n)}$. Thus, there exists a fixed binning with the corresponding pmf $\bar{p}$ such that if we replace $P$ with $\bar{p}$ in \eqref{eq:pmf} and denote the resulting pmf with $\hat{p}$, then $\hat{p}(y_1^n,y_2^n,g_{[0:2]})\apx{\epsilon^{(n)}}p^U(g_{[0:2]})p(y_1^n,y_2^n)$. Now, the second part of Lemma \ref{le:total} shows that there exists an instance $g_{[0:2]}$ such that $\hat{p}(y_1^n,y_2^n|g_{[0:2]})\apx{2\epsilon^{(n)}}p(y_1^n,y_2^n)$. Finally, eliminating $(\tR_0,\tR_1,\tR_2)$ from \eqref{eq:c1}, \eqref{eq:c2} and \eqref{eq:c3} by using Fourier-Motzkin elimination results in the rate region \eqref{eq:c0}.
\section{Converse }
 Let $Q$ denote a uniform random variable over $[1:n]$ and independent of all previously defined random variables. We choose single-letter
auxiliary random variables as follows: $U=(F_{2},B_{2},Y_{1:Q-1}^{(1)},Q)$  and
$V=(F_{1},B_{1},Y_{1:Q-1}^{(2)},Q)$. Using the fact that $I(B_{2};B_{1})=0$ that comes from the model (because $A_1$ and $A_2$ are creating these random variables at the beginning) we have:
\begin{align}
n(R_{f_{2}}+R_{f_{1}})&\ge H(F_{2})+H(F_{1})\nonumber\\
&\ge I(F_{2};F_{1}B_{1}\mid B_{2})+I(B_{2};F_{1}\mid B_{1})\nonumber\\
&=I(F_{2}B_{2};F_{1}B_{1})\nonumber\\
&\ge I(F_{2}B_{2};Y_{1}^{n})\nonumber\\
&\ge \sum_{q=1}^{n}I(F_{2}B_{2};Y_{q}^{(1)}\mid Y_{1:q-1}^{(1)})\nonumber\\
&=\sum_{q=1}^{n}[I(F_{2}B_{2}Y_{1:q-1}^{(1)};Y_{q}^{(1)})-I(Y_{1:q-1}^{(1)};Y_{q}^{(1)})]\nonumber\\
&\ge\sum_{q=1}^{n}I(F_{2}B_{2}Y_{1:q-1}^{(1)};Y_{q}^{(1)})-n g_{1}(\epsilon)\label{eqn:ae1}\\
&=nI(F_{2}B_{2}Y_{1:Q-1}^{(1)};Y_{Q}^{(1)}|Q)-n g_{1}(\epsilon),\nonumber\\
&\ge nI(F_{2}B_{2}Y_{1:Q-1}^{(1)},Q;Y_{Q}^{(1)})-n g_{1}(\epsilon)-ng_{2}(\epsilon) \label{eqn:ae2}\\
&=nI(U;Y_{Q}^{(1)})-n g_{1}(\epsilon)-n g_2(\epsilon),\label{eq:a4}
\end{align}
where $g_{i}(\epsilon)$ stands for functions that converge to zero as $\epsilon$ converges to zero. Equations \eqref{eqn:ae1} and \eqref{eqn:ae2} hold, due to Lemma 20 and Lemma 21 of \cite{cuff:thesis}. In the same way one can show that \begin{align}n(R_{f_{2}}+R_{f_{1}}))\ge nI(V;Z_{Q}^{(1)})-g_{1}(\epsilon)-g_2(\epsilon)\label{eq:b4}.\end{align} Next in a similar fashion we have
\begin{align}
n(R_{f_{1}}+R_{b_{1}})&\ge H(F_{1}B_{1})\nonumber\\
&\ge I(F_{1}B_{1};Y_{2}^{n}Y_{1}^{n})\nonumber\\
&=\sum_{q=1}^{n}I(F_{1}B_{1};Y_{q}^{(1)}Y_{q}^{(2)}\mid Y_{1:q-1}^{(1)}Y_{1:q-1}^{(2)})\nonumber\\
&=\sum_{q=1}^{n}[I(F_{1}B_{1}Y_{1:q-1}^{(1)}Y_{1:q-1}^{(2)};Y_{q}^{(1)}Y_{q}^{(2)})\nonumber\\
&-I(Y_{1:q-1}^{(1)}Y_{1:q-1}^{(2)};Y_{q}^{(1)}Y_{q}^{(2)}]\nonumber\\
&\ge\sum_{q=1}^{n}[I(F_{1}B_{1}Y_{1:q-1}^{(2)};Y_{q}^{(1)}Y_{q}^{(2)})-g_{3}(\epsilon)]\n
&\ge nI(V;Y_{Q}^{(2)}Y_{Q}^{(1)})-ng_{3}(\epsilon)-ng_{4}(\epsilon).\label{eq:c4}
\end{align}
A similar statement can be proved for $n(R_{f_{2}}+R_{b_{2}})$. 

In summary, we have proved that for every $\epsilon$, any achievable rate tuple must belong to the set $\mr_{\mathsf{out}, \epsilon}$ defined as the set of all tuples $(R_{f_1},R_{f_2},R_{b_1}, R_{b_2})$  such that there exists $p(u,v,y_1,y_2)\in T_{\mathsf{out}, \epsilon}$ for which $(R_{f_1},R_{f_2},R_{b_1}, R_{b_2})$ satisfies the inequalities \eqref{eq:a4}, \eqref{eq:b4} and \eqref{eq:c4} where $T_{\mathsf{out}, \epsilon}$ is the set of $p(u,v,y_1,y_2)$ satisfying the Markov relations as in the definition of $T_{\mathsf{out}}$ and
\[\tv{p(y_1, y_2)-q(y_1, y_2)}<\epsilon.\] The proof continues by showing that $\cap_{\epsilon>0}\mr_{\mathsf{out}, \epsilon}=\mr_\mathsf{out}$. Note that the cardinality bounds can be proved using the standard Fenchel extension of the Caratheodory theorem \cite{elgamal}. This completes the proof for the converse.

\end{document}